\documentclass[aps,prd,10pt,twocolumn,superscriptaddress,nofootinbib,showkeys,showpacs,altaffilletter]{revtex4-1}

\usepackage{graphicx}
\usepackage{dcolumn}
\usepackage{amssymb}
\usepackage{amsmath}
\usepackage{amsfonts}
\usepackage{amsbsy}
\usepackage{color}
\usepackage{rotating}
\usepackage[english]{babel}
\usepackage{hyperref}

\newcommand{\be}{\begin{equation}}
\newcommand{\ee}{\end{equation}}
\newcommand{\bea}{\begin{eqnarray}}
\newcommand{\eea}{\end{eqnarray}}

\begin{document}

\title{Statistical hierarchy of varying speed of light cosmologies}

\date{\today}

\author{Vincenzo Salzano}
\affiliation{Institute of Physics, University of Szczecin, Wielkopolska 15, 70-451 Szczecin, Poland}
\author{Mariusz P. D\c{a}browski}
\affiliation{Institute of Physics, University of Szczecin, Wielkopolska 15, 70-451 Szczecin, Poland}
\affiliation{National Centre for Nuclear Research, Andrzeja So{\l}tana 7, 05-400 Otwock, Poland}
\affiliation{Copernicus Center for Interdisciplinary Studies, S{\l}awkowska 17, 31-016 Krak{\'o}w, Poland}


\begin{abstract}

Variation of the speed of light is quite a debated issue in cosmology with some benefits, but also with some controversial concerns. Many approaches to develop a consistent varying speed of light (VSL) theory have been developed recently. Although  a lot of theoretical debate has sprout out about their feasibility and reliability, the most obvious and straightforward way to discriminate and check if such theories are really workable has been missed out or not fully employed. What is meant here is the comparison of these theories with observational data in a fully comprehensive way. In this paper we try to address this point i.e., by using the most updated cosmological probes, we test three different candidates for a VSL theory (Barrow \& Magueijo, Avelino \& Martins, and Moffat) signal. We consider many different ans\"{a}tze for both the functional form of $c(z)$ (which cannot be fixed by theoretical motivations) and for the dark energy dynamics, in order to have a clear global picture from which we extract the results. We compare these results using a reliable statistical tool such as the Bayesian Evidence. We find that the present cosmological data is perfectly compatible with any of these VSL scenarios, but in one case (Moffat model) we have a higher Bayesian Evidence ratio in favour of VSL than in the standard $c=$ constant $\Lambda$CDM scenario. Moreover, in such a scenario, the VSL signal can help to strengthen constraints on the spatial curvature (with indication toward an open universe), to clarify some properties of dark energy (exclusion of a cosmological constant at $2\sigma$ level) and is also falsifiable in the nearest future due to some peculiar issues which differentiate this model from the standard model. Finally, we have applied some priors which come from cosmology and, in particular, from information theory and gravitational thermodynamics. They put stronger constraints on the models under consideration, though still in favour of two of the Moffat's proposals. 

\end{abstract}

\keywords{cosmology, speed of light, alternative gravity}

\pacs{$98.80-k,98.80.Es,98.80.Cq, 04.50.Kd$}


\maketitle

\section{Introduction}
\label{sec:Introduction}

Varying (or dynamical) constants theories have become well established alternative theories of gravity. Beginning with the works of Weyl \cite{Weyl}, Eddington \cite{Eddington}, and especially Dirac \cite{Dirac1937}, the first fully quantitative framework for such kind of a theory - varying gravitational constant $G$ theory - was set by Brans and Dicke \cite{BD}. In such a theory $G$, which is a dimensional constant, was related to an average gravitational potential (scalar field) surrounding a given particle in the universe. It emerged contemporary that this $G$-dynamical theory became the key support for the low-energy-effective superstring common sector action \cite{superstring}, where the string coupling constant changes and the scalar field related to such a change became one of the superstring particle - the dilaton.

Despite the two canons of special relativity - the relativity principle and, in particular, the light principle - the theories of dynamical speed of light have already been considered by Einstein himself \cite{Einstein1907}, followed by Dicke \cite{Dicke1957}, Petit \cite{Petit1988}, and Moffat \cite{Moffat93A,Moffat93B}. These were followed by the VSL theories of Albrecht and Magueijo \cite{Albrecht99,Barrow99} in which Lorentz invariance is broken (relativity principle and light principle) so that there is a preferred frame in which the field is minimally coupled to gravity. Then, the geometrical tensors are computed in such a frame and no additional (boundary) terms appear in this frame (though they do in other frames) and the form of Einstein equations remains the same. There were also other proposals, e.g. Magueijo covariant (conformally) and locally invariant theory \cite{Magueijo00} which had as limits both minimal VSL theory and Brans-Dicke theory. Avelino and Martins \cite{Avelino99,Avelino00} have then  considered a small modification of the theory by Albrecht and Magueijo, and very recently Moffat proposed the dynamical $c$ theory with an extra vector field driving spontaneous violation of Lorentz invariance \cite{Moffat16}.

Different examples of VSL theories (though with $c$ basically depending on energy rather than time) can be considered within the framework of the so-called rainbow gravity theories which are based on the modified dispersion relations which deform Lorentz group to include Planck energy as a second invariant (Doubly Special Relativity - DSR) \cite{MagSmolin}.

The main objections to dynamical speed of light theories are that they refer to a dimensional constant rather than to a dimensionless one \cite{Duff}, and that the speed of light enters many apparently different physical theories (relativity, electrodynamics, field theory) and so any change in one of them should be consistent with all of them \cite{EllisUzan2005}. In this paper we will not be going into the discussion of these fundamentals (though appreciating them), trying instead to make some observational insight into the matter which we think is reasonable on the ground of physics which is an experimental science.

The discussion about the dimensionality of $c$ problem also refers to the dimensionality of $G$, which is rarely objected by physicists dealing with Brans-Dicke type of theories. We stay on a side that even if considering dimensional $c$, its dynamics can always be related to a dimensionless quantity such as the fine structure constant and so may become fully acceptable.  As it is well-known, varying fine structure constant theories have been studied intensively both in the theory \cite{alpha} and in observations \cite{alphaobs} among last two decades and apparently they started from the Bekenstein theory of varying electron charge $e$ model \cite{Bekenstein} which seems to be on the same footing as varying speed of light models due to the definition of the fine structure constant $\alpha = e^2/\hbar c$, where $\hbar$ is the Planck constant.

Thus, we are of course aware that there is a large debate about the benefit to study and develop VSL theories, i.e. based on the variation of dimensional physical constants. Here, we want to make clear our position and explain why we think that studies like the one performed in this work still have some importance.

We think one should consider separately the theoretical study and the effective possibility to test VSL theories. Even assuming that the only meaningful theories possible are those of dimensionless quantities, we need to stress out that theories involving the variation of the speed of light can indeed be made dimensionless. In fact, in some recent works \citep{PRL15,PRD16}, we have already focused on varying speed of light (VSL) theories, and have shown that they can leave a signature in the cosmological background (in particular, in the galaxy distribution) which can be measured through the quantity $D_{A}(z) H(z) / c_{0}$, where $D_{A}$ is the angular diameter distance, $H$ is the cosmological expansion rate, and $c_{0}$ is the speed of light measured here and now. In particular, the evaluation of such a quantity at the maximum redshift, defined as the redshift where $D_{A}$ reaches its maximum, is a strong discriminant: it would be equal to one only in the case of constant speed of light while, if there was any variation, it would be different from one. Some other observational tests for VSL cosmologies have also been suggested \cite{PLB14,JCAP14}. In those previous works, we have not questioned about the physical mechanism (and, thus, the theoretical background) behind a VSL signal; we have simply assumed that there is one, and we have tried to find out the way to detect it. In this work we  focus on a different subject: among many VSL theories/approaches, we want to analyse if there is a (statistical) hierarchy, i.e. if some approach is more or less compatible with observational data; if it is more or less probable (by means of some statistical tools) than a ``classical'' constant speed of light scenario. If such theories are falsifiable through measurements of dimensionless quantities, then, it would be a second step; and we will discuss further if this is possible.

The paper is organized as follows. In Sec.~\ref{sec:Models} we give the most important and general details about the VSL theories we analyse; in Sec.~\ref{sec:data} we describe the present observational status we use to constrain our models. In Sec.~\ref{sec:statistical} we discuss statistical tools and priors applied to VSL models while in Sec.~\ref{sec:results} we show all the main results from our statistical analysis. Finally, in Sec.~\ref{sec:conclusions} we put all the main conclusions we can derive from our work.

\section{Models}
\label{sec:Models}

In this section we discuss different VSL approaches. We describe the main properties of them while for further details, the interested reader is referred to the suggested bibliography. In analogy with standard notation, we introduce and define the following quantities
\begin{equation}
H_{c}(t) \equiv \frac{\dot{c}(t)}{c(t)} \; \qquad q_{c}(t) \equiv -\frac{\ddot{c}(t) c(t)}{\dot{c}^{2}(t)}\;,
\end{equation}
which can be understood as the kind of analogy to the Hubble parameter and the deceleration parameter for the varying speed of light dynamics.

\subsection{Barrow \& Magueijo (BM) model}

One of the first approaches to a reliable VSL was introduced and explored in a series of papers, \citep{Albrecht99,Barrow99,BarrowMagueijo99A,BarrowMagueijo99B,BarrowMagueijo99C,BarrowMagueijo00,Magueijo00}; we will call it BM approach. In a VSL theory, Lorentz invariance is broken and we have a preferred frame for the formulation of the physical laws. In such a frame, the speed of light is promoted to a field, $\Phi(x) = c^4(x)$, with a time variation included. Then, such a field is assumed to be minimally-coupled to gravity, with the action given by
\begin{equation}\label{eq:action_BM}
S = \int d ^{4} x \left[ \sqrt{-g} \left(\frac{\Phi (R+2\Lambda)}{16\pi G} + \mathcal{L}_{M} \right) + \mathcal{L}_{\Phi} \right]\; ,
\end{equation}
where: $g$ is the determinant of the metric $g_{\mu\nu}$; $R$ is the Ricci scalar; $\mathcal{L}_{M}$ is the Lagrangian of any considered matter component; $\mathcal{L}_{\Phi}$ is the Lagrangian of the field $\Phi$ and does not contain any metric contribution, in order to have a minimal coupling; $G$ is the universal gravitational constant (which we assume here to be constant, while in \citep{BarrowMagueijo99B} it is not); and $\Lambda$ is the cosmological constant (we introduce it here as in the original papers we are referring to, but we will consider a more general dark energy fluid in our analysis, not only a cosmological constant). Varying the action in the preferred frame, one finds \citep{Barrow99} that the Einstein's equations stay unchanged with respect to the classical $c$-constant case. Assuming the universe is homogeneous and isotropic, we end with the following Friedmann and acceleration equations,
\begin{equation}\label{eq:Friedmann_BM}
H^{2}(t) = \frac{8 \pi G}{3} \rho(t) - \frac{k \, c^{2}(t)}{a^{2}(t)} \; ,
\end{equation}
\begin{equation}\label{eq:acceleration_BM}
\frac{\ddot{a}(t)}{a(t)} = -\frac{4 \pi G}{3} \left( \rho(t) + 3 \frac{p(t)}{c^{2}(t)}\right)\; ,
\end{equation}
where $\rho(t)$ stands for the total matter budget in the universe and $k = 0, \pm 1$ for the spatial curvature. The interesting point is that, when they are combined in order to obtain the continuity equation, the presence of a time varying $c(t)$ leads to
\begin{equation}\label{eq:continuity_BM}
\dot{\rho}(t) + 3 H(t) \left( \rho(t) + \frac{p(t)}{c^{2}(t)} \right) = \frac{3 k \, c^{2}(t)}{4 \pi G \, a^{2}(t)} H_{c}(t)\; ,
\end{equation}
with a new term on the right hand side which is, in general, different from zero and brings, as expected, a violation of energy conservation. It is interesting to note that such term is related to both VSL and spatial curvature, with a possible degeneracy between them. Moreover, in a spatially flat universe, with $k=0$, we would have no way to determine if the speed of light $c$ is varying or not, because neither the fluid behaviour (through the continuity equation), nor the background dynamics (through the first Friedmann equation) would retain any possible signature to discriminate between a VSL signal and a constant speed of light.

\subsection{Avelino \& Martins (AM) model}

The first generalisation to the previous approach was given in \citep{Avelino99,Avelino00}. The authors assumed arbitrary variations of the speed of light, not restricted to the minimal coupling, thus leaving open the possibility for the VSL to contribute curvature corrections. This model will be called AM approach. In this case, the Friedmann and acceleration equations are
\begin{equation}\label{eq:Friedmann_AM}
H^{2}(t) = \frac{8 \pi G}{3} \rho(t) - \frac{k \, c^{2}(t)}{a^{2}(t)} \; ,
\end{equation}
\begin{equation}\label{eq:acceleration_AM}
\frac{\ddot{a}(t)}{a(t)} = -\frac{4 \pi G}{3} \left( \rho(t) + 3 \frac{p(t)}{c^{2}(t)}\right) + H(t) H_{c}(t)\; .
\end{equation}
Combining them, we obtain the continuity equation
\begin{equation}\label{eq:continuity_AM}
\dot{\rho}(t) + 3 H(t) \left( \rho(t) + \frac{p(t)}{c^{2}(t)} \right) = 2 \rho(t) H_{c}(t) \; .
\end{equation}
The main difference with the previous section is that a violation of the energy conservation is not directly related to the spatial curvature, e.g., the influence of the VSL theory is geometrically independent and represented by the last term on the right-hand side of Eq.~(\ref{eq:continuity_AM}).

\subsection{Moffat model}

The very first attempt to build a consistent VSL theory was given in \citep{Moffat93A,Moffat93B} and then elaborated in \citep{Clayton99,Clayton00,Clayton01,Clayton02}. Here, we will consider the most recent update from the same author, as discussed in \citep{Moffat16}, and called Moffat approach here. With respect to previous approaches, this latter model is much more complex, with many more ingredients and degrees of freedom. In some way, it might be considered as a more complete and general way in which a VSL might be introduced. In this model, the general action is made of up to four terms,
\begin{equation}\label{eq:action_Mof}
S = S_{G} + S_{\psi}  + S_{\phi} + S_{M}\; ,
\end{equation}
where: $S_{G}$ is the usual gravitational action, with the speed of light promoted to a field, $\Phi(x) = c^{4}(x)$, and no minimal coupling requirement is assumed,
\begin{equation}
\label{eq:Moffat}
S_{G} = \frac{1}{16\pi G} \int d ^{4} x \sqrt{-g}  \left[ \Phi (R+2\Lambda) -\frac{\kappa}{\Phi} \partial^{\sigma}\Phi \partial_{\sigma}\Phi \right]\; ,
\end{equation}
with $\kappa$ a dimensionless constant. $S_{\psi}$ is the action representing the dynamics of a vector field $\psi_{\mu}$ driving a spontaneous violation of $SO(3,1)$ Lorentz invariance, and is given by
\begin{equation}
S_{\psi} = \int d^{4}x  \sqrt{-g} \left[-\frac{1}{4} B^{\mu\nu}B_{\mu\nu} - W(\psi_{\mu}) \right]\; ,
\end{equation}
with $B_{\mu\nu} = \partial_{\mu} \psi_{\nu} -\partial_{\nu} \psi_{\mu}$, and $W$ a potential;
$S_{\phi}$ is the action of a dimensionless scalar field $\phi$, minimally coupled to gravity, and which should be responsible for quantum primordial fluctuations,
\begin{equation}
S_{\phi} = \frac{c^{4}_{0}}{16 \pi G}\int d ^{4} x  \sqrt{-g} \left[\frac{1}{2} (\partial_{\mu}\phi \partial^{\mu}\phi) - V(\phi) \right]\; ,
\end{equation}
with $V$ a potential; $S_{M}$ is the matter fields action, and $c_{0}$ the value of the speed of light today. Assuming isotropy and homogeneity, and dropping fields $\psi$ and $\phi$ as appropriate only to the very early stage of the universe expansion \citep{Moffat-PC}, we end up with the following Friedmann and acceleration equations:
\begin{equation}\label{eq:Friedmann_Mof}
H^{2}(t) = \frac{8 \pi G}{3} \rho(t) - \frac{k \, c^{2}(t)}{a^{2}(t)} - 4 H(t) H_{c}(t)
+ \frac{8 \kappa}{3} H^{2}_{c}(t)\; ,
\end{equation}
\begin{eqnarray}\label{eq:acceleration_Mof}
\frac{\ddot{a}(t)}{a(t)} &=& -\frac{4 \pi G}{3} \left( \rho(t) + 3 \frac{p(t)}{c^{2}(t)}\right) -2 H(t) H_{c}(t) \nonumber \\
&-& 2 \left(3 + \frac{8 \kappa}{3} -q_{c} \right) H^{2}_{c}(t).
\end{eqnarray}
Combining these two equations we find a modified continuity equation which looks as
\begin{eqnarray}\label{eq:continuity_Mof}
\dot{\rho}(t) &=& - 3 H(t) \left( \rho(t) + \frac{p(t)}{c^{2}(t)} \right) - 2 H_{c}(t) \left( \rho(t) + 3\frac{p(t)}{c^{2}(t)}\right) \nonumber \\
&+&\frac{3 k c^2(t)}{4\pi G a^2(t)} H_{c}(t) + \frac{4}{4\pi G}(3+2\kappa) H^{2}_{c}(t) \\
&\times& \left\{H_{c}(t)\left[q_{c}(t) -3 \right]-3 H(t)\right\} \; .\nonumber
\end{eqnarray}
If we take into account the equation of motion for the field $\Phi$, i.e.
\begin{equation}
\nabla^{\alpha}\nabla_{\alpha} \Phi = \frac{8\pi G}{3+2\kappa} T\;
\end{equation}
where $T$ is the trace of the energy-momentum tensor, and which can be found to be explicitly equivalent to
\begin{equation}
4 H^{2}_{c}(t) \left[ q_{c}(t)-3 \right] - 12 H(t)H_{c}(t) = \frac{8\pi G}{3+2\kappa} \left( \rho + 3\frac{p(t)}{c^{2}(t)} \right)
\end{equation}
then the continuity equation reduces to
\begin{equation}\label{eq:continuity_Mof_end}
\dot{\rho}(t) + 3 H(t) \left[ \rho(t) + \frac{p(t)}{c^{2}(t)} \right] = \frac{3 k c^{2}(t)}{4\pi G a^{2}(t)} H_{c}(t) \; .
\end{equation}
Thus, in a theoretical context where the VSL is properly considered as a field, we reach the same continuity equation from BM, but a Friedmann equation which is different from both BM and AM approaches. It is worth mentioning that the action (\ref{eq:Moffat}) is a special case of a general scalar-tensor action of the form \cite{Polarski}
\begin{equation}
S_{\Psi} = \frac{1}{16\pi G} \int d ^{4} x \sqrt{-g}  \left[ F(\Psi) R - Z(\Psi) \partial^{\sigma}\Psi \partial_{\sigma}\Psi - 2U(\Psi) \right]\; ,
\end{equation}
with an appropriate choice of the functions $F(\Psi)$ and $Z(\Psi)$ (see also \cite{EllisUzan2005}).

\subsection{VSL and DE ans\"{a}tze}

In order to apply the previous equations to cosmological data, we need some further elements to be specified. First of all, in all continuity equations we have to specify the equation of state (EoS) for all matter and energy fluids involved in the dynamics of the universe. Given the pressure as $p(t) = w(t) \rho(t) c^{2}(t)$, with $w(t)$ the EoS parameter, we will consider the following components in the cosmological fluid:
\begin{itemize}
   \item $w_{m} = 0$, for pressureless non-relativistic matter (both baryonic and dark matter);
   \item $w_{r} = 1/3$, for radiation (photons);
   \item $w_{DE} = -1$, for the cosmological constant case, when considering non-dynamical dark energy;
   \item $w_{DE} = w_{0} + w_{1} \, (1-a)$, the well known Chevallier-Polarski-Linder (CPL) model \citep{Chevallier01,Linder03}, mostly used as reference model for dynamical dark energy models.
\end{itemize}
Moreover, we further need to specify the $c(t)$ functions; we have considered three different ans\"{a}tze, i.e.
\begin{itemize}
   \item $c(a) = c_{0} \, a^{n}$ or $H_c = nH$ (named ``$c$-cl'' in tables), the classical and most general ansatz used, for example, in \citep{BarrowMagueijo99B};
   \item $c(a) = c_{0} \left[1 + \left(\frac{a}{a_{c}}\right)^{n} \right]$ (named ``$c$-Mag'' in tables), proposed in \citep{Magueijo00} and here applied for the first time to cosmological data. It describes a transition from some $c\neq c_{0}$ to $c_{0}$ at some epoch $a_{c}$; how fast will be the transition will be ruled by the parameter $n$;
   \item $c(a) = c_{0} \left[1 + n \, (1-a) \right]$ (named ``$c$-CPL'' in tables), a linear (in scale factor) VSL \textit{\`{a}-la CPL}.
\end{itemize}
In order to calculate cosmological distances through the explicit expression of $H(t)$ (or $H(a)$, with $a$ the cosmological scale factor) we need the Friedmann equation, and $H(t)$ will have to contain the explicit time-behaviour expression for matter, radiation and dark energy fluids, which can be derived solving the continuity equations. In the BM approach, we have already discussed how the energy violation is intrinsically related to the value of the spatial curvature; as we are going to consider the most general case of $k\neq0$, we can have non-trivial solutions to the continuity equation. We note that the right hand side of Eqs.~(\ref{eq:continuity_BM}) and (\ref{eq:continuity_Mof_end}) are not dependent explicitly on the density. Thus, we can assume that fluids evolve separately (i.e. there is no interaction among them), and we can write Eq.~(\ref{eq:continuity_BM}) as:
\begin{equation}
\rho'_{i}(a) +  \frac{3}{a} \left[ 1 + w_{i}(a) \right] \rho_{i}(a) = 0
\end{equation}
for matter and radation (labeled by index ``\textit{i}''), and as
\begin{equation}
\rho'_{DE}(a) +  \frac{3}{a} \left[ 1 + w_{DE}(a) \right] \rho_{DE}(a) =  \frac{3 k \, c^{2}(a)}{4 \pi G \, a^{2}} \frac{c'(a)}{c(a)}
\end{equation}
for dark energy (DE). We have also made use of the relation $\dot{a} = d a / d t$ in order to change the time derivatives (dots) with scale factor ones (primes); in this way, by solving the differential equation, we can express densities directly as functions of the scale factor. Then, in the BM and Moffat approaches, we are implicitly assuming that matter and radiation behave in a ``classical'' way, while the VSL signal is mainly influencing (and can contribute to) the dark energy fluid. Actually, to uncover how much the DE amount and trend will be changed (in a significant or not observationally detectable way) by the VSL assumption is another purpose of this work. Finally, it may be worth to note down that even in the case of a cosmological constant, i.e. $w_{DE} = -1$, we will not have a constant DE fluid but, instead, a ``dynamical'' cosmological constant, due to the $k$-$c'$ term.

In the AM approach, things are different: the VSL influence does not strictly depend on geometry, being always present. And it affects the continuity equation through a term which is proportional to the density. Thus, in this case, even matter and radiation will result to have a different dynamical behaviour with respect to the classical scenario,
\begin{equation}
\rho'_{i}(a) + \frac{3}{a} \left[ 1 + w_{i}(a) \right] \rho_{i}(a)  = 2 \rho_{i}(a) \frac{c'(a)}{c(a)} \; .
\end{equation}

The good point of these approaches is that, given the way their continuity equations are expressed, we can solve them analytically in most cases (only when using a CPL DE model we might have to solve them numerically for some $c(z)$-ans\"{a}tze), and express densities directly as functions of scale factor (or redshift), which turns out to be vital for the application to cosmological data. Some basic results were also given in the seminal works, here we have extended them to a more general records of VSL forms.

\section{Data}
\label{sec:data}

The analysis has involved the largest updated set of cosmological data available so far, and includes: expansion rate data from early-type galaxies; Type Ia Supernovae; Baryon Acoustic Oscillations; Cosmic Microwave Background; and a prior on the Hubble constant parameter, $H_{0}$. For the sake of consistency, in next subsections all the cosmological distances involved in our analysis will be expressed in the case of null spatial curvature, i.e. for $\Omega_{k} = 0$. But in our analysis, we have left the curvature free, so that they should be extended to the general-$\Omega_{k}$ case following comoving distance \citep{Hogg}
\begin{equation}\label{eq:DM_k}
D_{M}(z) = \begin{cases} \frac{D_{H}}{\sqrt{\Omega_{k}}} \sinh \left( \sqrt{\Omega_{k}} \frac{D_{C}(z)}{D_{H}} \right) &\mbox{for } \Omega_{k} > 0\\
D_{C}(z) &\mbox{for } \Omega_{k} = 0 \\
\frac{D_{H}}{\sqrt{|\Omega_{k}|}} \sin \left( \sqrt{|\Omega_{k}|} \frac{D_{C}(z)}{D_{H}} \right) &\mbox{for } \Omega_{k} < 0 \, ,
\end{cases}
\end{equation}
where $D_{H} = c_{0} / H_{0}$ is the Hubble distance, $D_{C}(z) = D_{H} \int^{z}_{0} dz' / E(z')$ is the line-of-sight comoving distance, $E(z) = H(z)/H_{0}$, $\Omega_k$ is the dimensionless curvature density parameter, and considering that luminosity distance and angular diameter distance are given respectively by
\begin{eqnarray}
D_{L}(z) &=& (1+z) D_{M}(z)\; , \\
D_{A}(z) &=& \frac{D_{M}(z)}{1+z} \; .
\end{eqnarray}
How to generalize all these equations to the case of a VSL model will be described in detail below, case by case.

\subsection{Hubble data from early-type galaxies}

We use a compilation of Hubble parameter measurements estimated with the differential evolution of passively evolving early-type galaxies as cosmic chronometers, in the redshift range $0<z<1.97$, and recently updated in \citep{Moresco15}. The corresponding $\chi^2_{H}$ estimator is defined as
\begin{equation}\label{eq:hubble_data}
\chi^2_{H}= \sum_{i=1}^{24} \frac{\left( H(z_{i},\boldsymbol{\theta_{b}},\boldsymbol{\theta_{c}})-H_{obs}(z_{i}) \right)^{2}}{\sigma^2_{H}(z_{i})} \; ,
\end{equation}
with $\sigma_{H}(z_{i})$ the observational errors on the measured $H_{obs}(z_{i})$ values, $\boldsymbol{\theta_{b}}$ the vector of the cosmological background parameters, and $\boldsymbol{\theta_{c}}$ the vector of the parameters directly related to the functional form of $c(z)$. Moreover, we will add a gaussian prior, derived from the Hubble constant value given in \citep{Bennett14}, $H_{0} = 69.6 \pm 0.7$.

\subsection{Type Ia Supernovae}

We used the SNeIa (Supernovae Type Ia) data from the JLA (Joint-Light-curve Analysis) compilation \citep{JLA}. This set is made of $740$ SNeIa obtained by the SDSS-II (Sloan Digital Sky Survey) and SNLS (Supenovae Legacy Survey) collaboration, covering a redshift range $0.01<z<1.39$. The $\chi^2_{SN}$ in this case is defined as
\begin{equation}
\chi^2_{SN} = \Delta \boldsymbol{\mu} \; \cdot \; \mathbf{C}^{-1}_{SN} \; \cdot \; \Delta  \boldsymbol{\mu} \; ,
\end{equation}
with $\Delta\boldsymbol{\mu} = \mu_{theo} - \mu_{obs}$, the difference between the observed and the theoretical value of the observable quantity for SNeIa, the distance modulus; and $\mathbf{C}_{SN}$ the total covariance matrix (for a discussion about all the terms involved in its derivation, see \citep{JLA}). The predicted distance modulus of the SNeIa, $\mu$, given the cosmological model and two other quantities, the stretch $X_{1}$ (a measure of the shape of the SNeIa light-curve) and the color $\mathcal{C}$, is defined as
\begin{equation}\label{eq:m_jla}
\mu(z,\boldsymbol{\theta}_{b},\boldsymbol{\theta}_{c}) = 5 \log_{10} [ D_{L}(z, \boldsymbol{\theta_{b}}, \boldsymbol{\theta_{c}}) ] - \alpha X_{1} + \beta \mathcal{C} + \mathcal{M}_{B} \; ,
\end{equation}
where $D_{L}$ is the luminosity distance. In a classical context, where $c$ is constant, this is given by
\begin{equation}\label{eq:dL}
D_{L}(z, \boldsymbol{\theta_{b}} )  = \frac{c_{0}}{H_{0}} (1+z) \int_{0}^{z} \frac{\mathrm{d}z'}{E(z',\boldsymbol{\theta_{b}})} \; ,
\end{equation}
with $H(z) \equiv H_{0} E(z)$ (following \citep{JLA}, only for SNeIa analysis we assume $H_{0} = 70$ km/s Mpc$^{-1}$) and $c_{0}$ the speed of light measured here and now. The vector $\boldsymbol{\theta_{b}}$ will include cosmologically-related parameters and three other fitting parameters: $\alpha$ and $\beta$, which characterize the stretch-luminosity and color-luminosity relationships; and the nuisance parameter $\mathcal{M}_{B}$, expressed as a step function of two more parameters, $\mathcal{M}^{1}_{B}$ and $\Delta_{m}$:
\begin{equation}
\mathcal{M}_{B} = \begin{cases} \mathcal{M}^{1}_{B} & \mbox{if} \quad M_{stellar} < 10^{10} M_{\odot}, \\
\mathcal{M}^{1}_{B} + \Delta_{m} & \mbox{otherwise}.
\end{cases}
\end{equation}
Further details about this choice are given in Ref. \citep{JLA}.

When $c$ is varying according to our models, Eq.~(\ref{eq:dL}) is modified into
\begin{equation}\label{eq:mu_jla}
D_{L}(z, \boldsymbol{\theta_{b}}, \boldsymbol{\theta_{c}} )  = \frac{1}{H_{0}} (1+z) \int_{0}^{z} \frac{c(z',\boldsymbol{\theta_{c}})}{E(z',\boldsymbol{\theta_{b}})} \mathrm{d}z'\; .
\end{equation}

\subsection{Baryon Acoustic Oscillations}

The $\chi^2_{BAO}$ for Baryon Acoustic Oscillations (BAO) is defined as
\begin{equation}
\chi^2_{BAO} = \Delta \boldsymbol{\mathcal{F}}^{BAO} \; \cdot \; \mathbf{C}^{-1}_{BAO} \; \cdot \; \Delta  \boldsymbol{\mathcal{F}}^{BAO} \; ,
\end{equation}
where the quantity $\mathcal{F}^{BAO}$ can be different depending on the considered survey. We used data from the WiggleZ Dark Energy Survey, evaluated at redshifts $z=\{0.44,0.6,0.73\}$, and given in Table~1 of \citep{WiggleZ}; in this case the quantities to be considered, when $c$ is constant, are the acoustic parameter
\begin{equation}\label{eq:AWiggle}
A(z, \boldsymbol{\theta_{b}}) = 100  \sqrt{\Omega_{m} \, h^2} \frac{D_{V}(z,\boldsymbol{\theta_{b}})}{c_{0} \, z} \, ,
\end{equation}
and the Alcock-Paczynski distortion parameter
\begin{equation}\label{eq:FWiggle}
F(z, \boldsymbol{\theta_{b}}) = (1+z)  \frac{D_{A}(z,\boldsymbol{\theta_{b}})\, H(z,\boldsymbol{\theta_{b}})}{c_{0}} \, ,
\end{equation}
where $D_{A}$ is the angular diameter distance
\begin{equation}\label{eq:dA}
D_{A}(z, \boldsymbol{\theta_{b}} )  = \frac{c_{0}}{H_{0}} \frac{1}{1+z} \ \int_{0}^{z} \frac{\mathrm{d}z'}{E(z',\boldsymbol{\theta_{b}})} \; ,
\end{equation}
and $D_{V}$ is the geometric mean of the physical angular diameter distance $D_A$ and of the Hubble function $H(z)$, and defined as
\begin{equation}\label{eq:dV}
D_{V}(z, \boldsymbol{\theta_{b}} )  = \left[ (1+z)^2 D^{2}_{A}(z,\boldsymbol{\theta_{b}}) \frac{c_{0} \, z}{H(z,\boldsymbol{\theta_{b}})}\right]^{1/3}.
\end{equation}
When dealing with varying $c$, Eqs.~(\ref{eq:AWiggle})-(\ref{eq:dA}) and (\ref{eq:dV}) have to be changed into
\begin{equation}\label{eq:AWiggle2}
A(z, \boldsymbol{\theta_{b}}, \boldsymbol{\theta_{c}}) = 100  \sqrt{\Omega_{m} \, h^2} \frac{D_{V}(z,\boldsymbol{\theta_{b}},\boldsymbol{\theta_{c}})}{c(z, \boldsymbol{\theta_{c}}) \, z} \, ,
\end{equation}
\begin{equation}\label{eq:FWiggle2}
F(z, \boldsymbol{\theta_{b}}, \boldsymbol{\theta_{c}}) = (1+z)  \frac{D_{A}(z, \boldsymbol{\theta_{b}},\boldsymbol{\theta_{c}})\, H(z, \boldsymbol{\theta_{b}},\boldsymbol{\theta_{c}})}{c(z, \boldsymbol{\theta_{c}})} \, ,
\end{equation}
\begin{equation}\label{eq:dA2}
D_{A}(z, \boldsymbol{\theta_{b}}, \boldsymbol{\theta_{c}} )  = \frac{1}{H_{0}} \frac{1}{1+z} \int_{0}^{z} \frac{c(z', \boldsymbol{\theta_{c}})}{E(z', \boldsymbol{\theta_{b}},\boldsymbol{\theta_{c}})} \mathrm{d}z'\; ,
\end{equation}
\begin{equation}\label{eq:dV2}
D_{V}(z, \boldsymbol{\theta_{b}}, \boldsymbol{\theta_{c}} )  = \left[ (1+z)^2 D^{2}_{A}(z, \boldsymbol{\theta_{b}},\boldsymbol{\theta_{c}}) \frac{c(z, \boldsymbol{\theta_{c}}) \, z}{H(z, \boldsymbol{\theta_{b}},\boldsymbol{\theta_{c}})}\right]^{1/3}.
\end{equation}
We have also considered the data from the SDSS-III Baryon Oscillation Spectroscopic Survey (BOSS) DR$12$, described in \citep{SDSS12} and expressed as
\begin{equation}
D_{M}(z) \frac{r^{fid}_{s}(z_{d})}{r_{s}(z_{d})} \qquad \mathrm{and} \qquad H(z) \frac{r_{s}(z_{d})}{r^{fid}_{s}(z_{d})} \, ,
\end{equation}
where $r_{s}(z_{d})$ is the sound horizon evaluated at the dragging redshift $z_{d}$; and $r^{fid}_{s}(z_{d})$ is the same sound horizon but calculated for a given fiducial cosmological model used, being equal to $147.78$ Mpc \citep{SDSS12}. The redshift of the drag epoch is well approximated by \citep{Eisenstein}
\begin{equation}\label{eq:zdrag}
z_{d} = \frac{1291 (\Omega_{m} \, h^2)^{0.251}}{1+0.659(\Omega_{m} \, h^2)^{0.828}} \left[ 1+ b_{1} (\Omega_{b} \, h^2)^{b2}\right]\; ,
\end{equation}
where
\begin{eqnarray}
b_{1} &=& 0.313 (\Omega_{m} \, h^2)^{-0.419} \left[ 1+0.607 (\Omega_{m} \, h^2)^{0.6748}\right], \nonumber \\
b_{2} &=& 0.238 (\Omega_{m} \, h^2)^{0.223}.
\end{eqnarray}
The sound horizon, in the classical context of constant $c$, is defined as:
\begin{equation}\label{eq:soundhor}
r_{s}(z, \boldsymbol{\theta_{b}}) = \int^{\infty}_{z} \frac{c_{s}(z')}{H(z',\boldsymbol{\theta_{b}})} \mathrm{d}z'\, ,
\end{equation}
with the sound speed
\begin{equation}\label{eq:soundspeed}
c_{s}(z) = \frac{c_{0}}{\sqrt{3(1+\overline{R}_{b}\, (1+z)^{-1})}} \; ,
\end{equation}
and
\begin{equation}
\overline{R}_{b} = 31500 \Omega_{b} \, h^{2} \left( T_{CMB}/ 2.7 \right)^{-4}\; ,
\end{equation}
with $T_{CMB} = 2.726$ K. Finally, we have also added data points from Quasar-Lyman $\alpha$ Forest from SDSS-III BOSS DR$11$ \citep{Lyman}:
\begin{eqnarray}
\frac{D_{A}(z=2.36)}{r_{s}(z_{d})} &=& 10.8 \pm 0.4\; , \\
\frac{c_{0}}{H(z=2.36) r_{s}(z_{d})}  &=& 9.0 \pm 0.3\; .
\end{eqnarray}
When working with varying-$c$ models, of course, we will have to change $D_{A}$ and $D_{V}$ as described above, and also the sound horizon, through the definition of the sound speed, Eq.~(\ref{eq:soundspeed}), which now will be
\begin{equation}
c_{s}(z, \boldsymbol{\theta_{c}}) = \frac{c(z, \boldsymbol{\theta_{c}})}{\sqrt{3(1+\overline{R}_{b}\, (1+z)^{-1})}}.
\end{equation}
Thus, we will have three different contributions to $\chi^{2}_{BAO}$, e.g., $\chi^{2}_{WiggleZ},\chi^{2}_{BOSS},\chi^{2}_{Lyman}$, depending on the data sets we consider.

\subsection{Cosmic Microwave Background}

The $\chi^2_{CMB}$ for Cosmic Microwave Background (CMB) is defined as
\begin{equation}
\chi^2_{CMB} = \Delta \boldsymbol{\mathcal{F}}^{CMB} \; \cdot \; \mathbf{C}^{-1}_{CMB} \; \cdot \; \Delta  \boldsymbol{\mathcal{F}}^{CMB} \; ,
\end{equation}
where $\mathcal{F}^{CMB}$ is a vector of quantities taken from \citep{WangWang}, where \textit{Planck} $2015$ data release is analyzed in order to give a set of quantities which efficiently summarize the information contained in the full power spectrum (at least, for the cosmological background, see \citep{WangMukherjee}), and can thus be used in alternative to the latter. In the classical context, the quantities are the CMB shift parameters:
\begin{eqnarray}
R(\boldsymbol{\theta_{b}}) &\equiv& \sqrt{\Omega_m H^2_{0}} \frac{r(z_{\ast},\boldsymbol{\theta_{b}})}{c_{0}} \nonumber \\
l_{a}(\boldsymbol{\theta_{b}}) &\equiv& \pi \frac{r(z_{\ast},\boldsymbol{\theta_{b}})}{r_{s}(z_{\ast},\boldsymbol{\theta_{b}})}\, ,
\end{eqnarray}
and the baryonic density parameter, $\Omega_b \, h^{2}$. Again, $r_{s}$ is the comoving sound horizon, evaluated at the photon-decoupling redshift $z_{\ast}$, given by the fitting formula \citep{Hu}:
\begin{equation}{\label{eq:zdecoupl}}
z_{\ast} = 1048 \left[ 1 + 0.00124 (\Omega_{b} h^{2})^{-0.738}\right] \left(1+g_{1} (\Omega_{m} h^{2})^{g_{2}} \right) \, ,
\end{equation}
with
\begin{eqnarray}
g_{1} &=& \frac{0.0783 (\Omega_{b} h^{2})^{-0.238}}{1+39.5(\Omega_{b} h^{2})^{-0.763}}\; , \\
g_{2} &=& \frac{0.560}{1+21.1(\Omega_{b} h^{2})^{1.81}} \, ;
\end{eqnarray}
while $r$ is the comoving distance defined as:
\begin{equation}
r(z, \boldsymbol{\theta_{b}} )  = \frac{c_{0}}{H_{0}} \int_{0}^{z} \frac{\mathrm{d}z'}{E(z',\boldsymbol{\theta_{b}})} \mathrm{d}z'\; .
\end{equation}
When considering varying-$c$ models, again, the sound horizon will change as described above, and the comoving distance will be
\begin{equation}
r(z, \boldsymbol{\theta_{b}}, \boldsymbol{\theta_{c}} )  = \frac{1}{H_{0}} \int_{0}^{z} \frac{c(z', \boldsymbol{\theta_{c}})}{E(z', \boldsymbol{\theta_{b}},\boldsymbol{\theta_{c}})} \mathrm{d}z'\; ,
\end{equation}
and the shift parameter $R$ will become
\begin{equation}
R(\boldsymbol{\theta_{b}},\boldsymbol{\theta_{c}}) \equiv \sqrt{\Omega_m H^2_{0}} \frac{r(z_{\ast}, \boldsymbol{\theta_{b}},\boldsymbol{\theta_{c}})}{c(z, \boldsymbol{\theta_{c}})} .
\end{equation}

Thus, the total $\chi^2_{Tot}$ will be the sum of: $\chi^{2}_{H_{0}}$, $\chi^{2}_{H}$, $\chi^{2}_{SN}$, $\chi^{2}_{WiggleZ}$, $\chi^{2}_{BOSS}$, $\chi^{2}_{Lyman}$ and $\chi^{2}_{CMB}$. We minimize $\chi^2_{Tot}$ using the Markov Chain Monte Carlo (MCMC) method. The cosmological parameters vector $\boldsymbol{\theta_{b}}$ will be equal to $\{\Omega_{m}, \Omega_{b}, \Omega_{k}, h, w_{DE}, \alpha, \beta, M^{1}_{B}, \Delta_{M}\}$, where $w_{DE}$ can be $-1$ when considering the cosmological constant case, or $\{w_{0},w_{1}\}$ when dealing with the CPL parametrization. Instead, the VSL parameters vector $\boldsymbol{\theta_{c}}$ will be made of $\{n, a_{c}\}$ (plus $\kappa$, the kinetic term for the VSL field, in the Moffat approach) depending on the chosen ansatz for $c(z)$. The parameter $h$ is defined in a standard way by $H_{0} \equiv 100 \, h$. The density parameters entering $H(z)$ are $\Omega_{m},\Omega_{r},\Omega_{k}$ with $\Omega_{DE}$ consequently in order to ensure the condition $E(z=0)=1$. Moreover, the radiation density parameter $\Omega_{r}$ will be defined \citep{WMAP} as the sum of photons and relativistic neutrinos
\begin{equation}
\Omega_{r} = \Omega_{\gamma} (1+0.2271 \mathcal{N}_{eff})\, ,
\end{equation}
where $\Omega_{\gamma} = 2.469 \times 10^{-5} \, h^{-2}$ for $T_{CMB}= 2.726$ K; and the number of relativistic neutrinos is assumed to be $\mathcal{N}_{eff} = 3.046$.

\section{Statistical comparison tools}
\label{sec:statistical}

If we plan to have a statistical hierarchy of VSL approaches, we need a statistical tool to define the reliability of one model against the other. There is plenty of tools which might be used, but each of them is plagued by some problems and that makes the choice of the right one tricky. A list of the most used in cosmology can be found in \citep{InfoCriteria}, including: the Akaike Information Criterion (AIC), in its original form \citep{AIC1}, and the corrected one \citep{AIC2}; the Residual Information Criterion (RIC), both original and extended \citep{RIC}; the Bayesian Information Criterion (BIC) \citep{BIC}; and the Deviance Information Criterion (DIC) \citep{DIC}. Independently of the way they are derived, they all look like a ``variation on a theme'', being equal (except for DIC) to the minimum in the $\chi^2$ plus some term which involves a combination of the number of theoretical parameters and of the number of data used in the analysis. Thus, they will always tend to penalize (even if in different measure) models with a higher number of parameters in a clearly biased way. Some of them have also intrinsic problems, for example: the ``dimensional inconsistency'' of AIC; or BIC-based inference being suspicious when some theoretical parameters are degenerate (which often happens when dealing with DE models or parameterizations); and they do not generally help to state a ``validity hierarchy'', instead leading to conflicting results (see the case analysed in \citep{InfoCriteria}, for example).

\subsection{Bayesian evidence}

Finally, it is generally recognized that the best way to statistically compare models is by means of the Bayesian Evidence for it. In this work, the Bayesian Evidence for each model is calculated using the algorithm described in \citep{Mukherjee06}. As this algorithm is stochastic, in order to take into account possible statistical noise, we run it $\sim 100$ times obtaining a distribution of values from which we extract the best value of the evidence as the median of the distribution. The Evidence, $\mathcal{E}$, is defined as the probability of the data $D$ given the model $M$ with a set of parameters $\boldsymbol{\theta}$, $\mathcal{E}(M) = \int\ \mathrm{d}\boldsymbol{\theta}\ \mathcal{L}(D|\boldsymbol{\theta},M)\ \pi(\boldsymbol{\theta}|M)$: $\pi(\boldsymbol{\theta}|M)$ is the prior on the set of parameters, normalized to unity, and $\mathcal{L}(D|\boldsymbol{\theta},M)$ is the likelihood function. As commented above, there are many other tools to compare models, but the Bayesian Evidence is considered the most reliable, even if it is not completely immune to problems, like its dependence on the choice of priors \citep{Nesseris13}. In order to minimize such problems, we have always used the same uninformative flat priors on the parameters, and over sufficiently wide ranges, so that further increasing them will have no impact on the results. More details about them are in next subsections.

Once the Bayesian Evidence is calculated, one can obtain the Bayes Factor, defined as the ratio of evidences of two models, $M_{i}$ and $M_{j}$, $\mathcal{B}^{i}_{j} = \mathcal{E}_{i}/\mathcal{E}_{j}$. If $\mathcal{B}^{i}_{j} > 1$,  model $M_i$ is preferred over $M_j$, given the data. We have used, separately, the cosmological constant and the CPL model, both with constant speed of light and null spatial curvature, as reference models $M_j$.

Even if the Bayes Factor $\mathcal{B}^{i}_{j} > 1$, one is not able yet to state how much better is model $M_i$ with respect to model $M_j$. For this, we choose the widely-used Jeffreys' Scale \citep{Jeffreys98}. In general, Jeffreys' Scale states that: if $\ln \mathcal{B}^{i}_{j} < 1$, the evidence in favor of model $M_i$ is not significant; if $1 < \ln \mathcal{B}^{i}_{j} < 2.5$, the evidence is substantial; if $2.5 < \ln \mathcal{B}^{i}_{j} < 5$, is strong; if $\mathcal{B}^{i}_{j} > 5$, is decisive. Negative values of $\ln \mathcal{B}^{i}_{j}$ can be easily interpreted as evidence against model $M_i$ (or in favor of model $M_j$). In \citep{Nesseris13}, it is shown that the Jeffreys' scale is not a fully-reliable tool for model comparison, but at the same time the statistical validity of the Bayes factor as an efficient model-comparison tool is not questioned: a Bayes factor $\mathcal{B}^{i}_{j}>1$ unequivocally states that the model $i$ is more likely than model $j$. We present results in both contexts for reader's interpretation.

\subsection{Priors}

The most straightforward priors are derived from general cosmological assumptions. In particular:
\begin{itemize}
  \item $0 < \Omega_{b} < \Omega_{m}$; no upper limit is given, because we leave $\Omega_{k}$ free;
  \item $\Omega_{DE}> 0$ or, more generally, $\rho_{DE}(a)>0$ for any scale factor in the range $[0,1]$. This sounds like a quite natural and obvious condition (matter/energy density is positive by definition) but it is not so trivial in our analysis. In fact, when solving the continuity equations, Eqs.~(\ref{eq:continuity_BM})~-~(\ref{eq:continuity_AM}) and (\ref{eq:continuity_Mof_end}), it can happen that, for some combination of the parameters $\boldsymbol{\theta_{b}}$, $\boldsymbol{\theta_{c}}$  (in particular, $\Omega_{k}$, $w_{DE}$, $n$ and $\kappa$), we might have $\rho_{DE}(a)<0$ over some redshift range;
  \item for the CPL case, $w_{0}+w_{a}<1$.
\end{itemize}
We will define such priors as the \textit{cosmological priors}. But constraints of different nature might be assumed. In fact, in the past, VSL have been analyzed in the context of their thermodynamical consequences \citep{Chimento2001,Youm2002}. For the sake of completeness, we will also discuss such priors in our analysis, but we want to make it clear that the only reliable results are, in our opinion, only those derived from applying the cosmological conditions described above. That is why there are many possible flaws behind the above-mentioned thermodynamical considerations.

The first point, referring to \citep{Chimento2001}, introduces what we could define as an \textit{information prior}. In that work the authors state in a simple way a possible relation among VSL and variation in the entropy of the universe. In particular, they state that \textit{``an increase in c means a widening of the past light cone of the observers. Automatically they acquire more information and the entropy decreases accordingly.''} Such affirmation is then used to state that the only possible VSL signal needs to have $\dot{c}<0$. This condition is what we will call {\it information prior} later on. It is easy to check that, given our ans\"{a}tze for the functional form of $c(z)$, it corresponds to $n<0$ for the $c$-cl and $c$-Mag cases, and $n>0$ for the $c$-CPL one. Actually, this condition is very restrictive and still disputable. In fact, the past light cone of the observers only defines the visible universe, a smaller (not closed) region of the entire universe. It seems that the information problem in this case is quite different from the information paradox for black holes, because, in our case, there is actually no proper loss or gain of information: the information in the entire universe is always the same, independently of what we observe or not.

The second approach is described in Ref. \citep{Youm2002}. Assuming the universe satisfies the first law of thermodynamics, in a VSL scenario the time variation of the entropy is found to be:
\begin{equation}\label{eq:youm}
\dot{s} = \left( 2 \rho - \frac{c^2}{4\pi G}\Lambda + \frac{3 k c^{2}}{4 \pi G a^{2}} \right) \frac{a^{3} c \dot{c}}{T}\; ,
\end{equation}
where $s$ is the entropy, $\rho$ is the total matter density, $\Lambda$ is the cosmological constant, and $T$ is the temperature of the Universe. This formula is more complete than its equivalent in \citep{Chimento2001}, because it takes into account a cosmological constant, and not only ordinary matter. Moreover, it exhibits a coupling between the cosmological constant and the VSL signal because the former is introduced as a geometrical component. In our approach, instead, we consider dark energy, whatever it is, as a fluid, thus contributing to the stress-energy tensor. As such, following the same step in \citep{Youm2002}, our equivalent expression for Eq.~(\ref{eq:youm}) will be:
\begin{equation} \label{eq:entropy_AV}
\dot{s} = 4 \rho \frac{a^{3} c \dot{c}}{T}\; ,
\end{equation}
for the Avelino \& Martins model, and:
\begin{equation} \label{eq:entropy_MM}
\dot{s} = \left( 2 \rho + \frac{3 k c^{2}}{4 \pi G a^{2}} \right) \frac{a^{3} c \dot{c}}{T}\; ,
\end{equation}
for the Barrow \& Magueijo and Moffat models. They give different results because of the different continuity equations, Eqs.~(\ref{eq:continuity_BM})~-~(\ref{eq:continuity_AM}) and (\ref{eq:continuity_Mof_end}). Now, $\rho$ includes both matter (dark and baryonic) and dark energy. Finally, Eq.~(\ref{eq:entropy_MM}) can be written in terms of observational parameters as:
\begin{equation} \label{eq:entropy_obs}
\dot{s} = \left[ 1 - \Omega_{k} \left( 1 + \frac{\Delta^{2}_{c}}{a^2} \right) \right] \frac{3 H^{2}}{4 \pi G}\frac{a^{3} c \dot{c}}{T}\; ,
\end{equation}
where $\Delta_{c} = c(z)/c_{0}$. 

It is then straightforward to find out that, for the Avelino \& Martins scenario, you can only have $\dot{c}>0$ in order to ensure $\dot{s}>0$; and this conclusion contradicts the information prior stated in \citep{Chimento2001}. On the other hand, in the case of Barrow \& Magueijo and Moffat models, you can still have either $\dot{c}<0$ or $\dot{c}>0$, and $\dot{s}>0$. When we  consider the condition $\dot{s}>0$, as derived from Eqs.~(\ref{eq:entropy_AV})~-~(\ref{eq:entropy_obs}), we will assume that an \textit{entropy prior} is being applied.

However, we have to stress that Eqs.~(\ref{eq:entropy_AV})~-~(\ref{eq:entropy_obs}) are derived assuming that the energy (or the number particles) is conserved in the universe; but VSL intrinsically violate this condition \citep{Magueijo00}, as it is also possible to check from the same continuity equations we have derived in this work, Eqs.~(\ref{eq:continuity_BM})~-~(\ref{eq:continuity_AM}) and (\ref{eq:continuity_Mof_end}). Thus, in principle, they should/could not be applied to VSL in a healthy way, and they could lead to conflicts and/or inconsistencies with the same VSL approaches (as it is shown in Table~\ref{tab:conditions}, which we will discuss in next sections).

\section{Results and Discussion}
\label{sec:results}

We will discuss the results obtained from applying different kinds of priors in separate sections, enlightening pros and cons for each of them. In Table~\ref{tab:conditions} we report all the priors we can apply, and show if they are compatible with each other. A ``\checkmark'' implies that a fit with the assumed conditions is possible, and gives reliable results. A ``$\dag$'' means that the prior conditions can be satisfied by some set of parameters values, but no reliable fit is possible. For example, we can have high values of the $\chi^2$ ($\sim 10^4-10^6$), implying an unsatisfactory fit of the observational data; and not physical cosmological parameters, as $h\rightarrow 1$, $\Omega_m \rightarrow 0$, or $w_0 \rightarrow 0$. Finally, a ``$\ddag$'' means that there is no set of parameters at all which can satisfy the considered priors.

\subsection{Cosmological priors}

In this case, all results for our analysis are summarized in Table~\ref{tab:results_LCDM} for the cosmological constant case, and in Table~\ref{tab:results_CPL} for the CPL case. As we have stated in the previous section, these will be considered by us as the main result of this work.

Looking at the values of the logarithm of the Bayes Factor for the cosmological constant case, we can see how the values are mostly $<|1|$ which means no strong statistical evidence in favor of one model in particular, were it the reference scenario ($c-$constant, with or without spatial null curvature) or one of the VSL one. Most of the cases are completely equivalent among each other, not only when relying on the values of the cosmological parameters coming out our statistical analysis, but also when looking to possible alternative smoking guns as, for example, the maximum distance from the galaxy distribution we discussed (see column $8$ in Table~\ref{tab:results_LCDM} and column $10$ in Table~\ref{tab:results_CPL}).

The AM approach, while being $<|1|$, show negative Bayes Factor, thus, it should be slightly disfavoured with respect to the $c$-constant case (and null spatial curvature). On the contrary, the BM and Moffat scenarios report $>1$ values for the Bayes Factor in the case of a linear VSL signal, pointing toward a substantial evidence in favor of the VSL scenarios against the classical one with no spatial curvature. It is also interesting to note that mostly all the VSL scenarios have a higher Bayes Factor than the $\Omega_{k}$-free classical scenario, which means they are in some way more favoured with respect to this model.

Focussing on the values of the cosmological parameters, it is clear that only the AM approach results in different values for $\Omega_m$ and $\Omega_k$, both much higher (and with larger errors) then the standard model. But, as stated above, this approach is statistically not favoured. On the other hand, the models with the highest Bayes Factors are perfectly consistent with the $\Lambda$CDM expectations, looking almost identical, even when the value of the maximum redshift is taken into account.

Moving to the VSL ans\"{a}tze, the first point to be stressed is how the c-Mag ansatz has an important flaw: the scale factor at which there should be transition from one $c\neq c_{0}$ to $c = c_{0}$ is basically unconstrained, could it range from $0$ to $1$ at (slightly less than) $2\sigma$ level. An interesting point is that both the BM and Moffat scenarios with the highest Bayes Factor, have an index $n$ (parameterizing the VSL signal and its deviation from the $c$-constant case, holding for $n=0$) which is consistent with $0$ only at $2\sigma$ confidence level. Such VSL signals also seem to partially smear the effect from the spatial curvature: while the $\Omega_k$ best fit values look to be the same, there is a sensible decrease in their error estimations, pointing toward a better assessment of an open Universe $(\Omega_{k}>0)$.

When moving to the CPL case, in general, results are the same, qualitatively speaking, even if now, as expected, the presence of a dark energy component increase the degeneracy among parameters. The AM approach results to work better than the cosmological constant case but, even if the Bayes Factors are all $>0$, still the VSL models have smaller Evidence than the classical scenario with non-null curvature (i.e. CPL plus $\Omega_{k}$ free). Again, the BM and Moffat approaches give better statistical performances, with $>1$ Bayes Factors. Now the curvature is consistent with null value; but the $c-$running parameters $n$ are not consistent with zero values at least at $2\sigma$ level. Moreover, in this case we have a clear difference in both the cosmological parameters outputs (for the Moffat case, we have less matter required) and in the maximum redshift for $D_{A}$, which could be a good observational probe to discriminate between the standard and the two alternative scenarios.

Finally, the most favoured statistical scenarios point clearly toward a phantom dark energy ($w_{DE}<-1$ at late times), and the cosmological constant case is excluded at least at $2\sigma$ level (see values for the $w_{1}$ parameter).

We will also give a further detail about such models: as explained in the Introduction, while having a Bayesian hierarchy of VSL approach is an initial step, a more fundamental and important one would be to have a good observational discriminator among them. And, possibly and preferably, such discriminator should be related to a dimensionless quantity. In our previous works \citep{PRL15,PRD16}, we have shown that this is possible: one has to calculate where is located (in terms of redshift) the maximum in the angular diameter distance, and then calculate the quantity $D_{A}(z_{M})H(z_{M})/c_{0}$ which, for a general VSL theory with free curvature $\Omega_{k}$, is
\begin{equation}
\frac{D_{A}(z_{M}) H(z_{M})}{c_{0}} = \Delta_{c}(z_{M}) \cdot \Delta_{k}(z_{M})\; ,
\end{equation}
where $z_{M}$ is the redshift at which the angular diameter distance reaches its maximum, and $\Delta_{c}(z_{M})$ and $\Delta_{k}(z_{M})$ are defined as
\begin{equation}
\Delta_{c}(z_{M}) = \frac{c(z_{M})}{c_{0}} \; ,
\end{equation}
and
\begin{equation}
\Delta_{k}(z_{M}) = \begin{cases}
\cosh \left( \sqrt{\Omega_{k}} \frac{D_{C}(z_{M})}{D_{H}} \right) &\mbox{for } \Omega_{k} > 0\\
1 &\mbox{for } \Omega_{k} = 0 \\
\cos \left( \sqrt{|\Omega_{k}|} \frac{D_{C}(z_{M})}{D_{H}} \right) &\mbox{for } \Omega_{k} < 0 \, .
\end{cases}
\end{equation}
$\Delta_{c}$ is, basically, the contribution to the signal given only by the relative variation of the speed of light between now and the $z_{M}$ epoch, while $\Delta_{k}$ quantifies the contribution to the signal from the spatial curvature. Actually, in $\Delta_{k}$ there is still some influence from $c(z)$, through the comoving distance $D_{C}$.

Thus, for each model we have considered, we have calculated the maximum redshift $z_{M}$, the total $\Delta_{c}\Delta_{k}$ signal, and the separate contributions from both VSL and curvature. In \citep{PRL15,PRD16} we have stated that, in principle, the Square Kilometer Array (SKA) \citep{SKA_site} will be able to detect, if any, a total signal\footnote{In \citep{PRL15,PRD16} we have assumed that such signal is made only by VSL, while the curvature is null, i.e., $\Delta_{k} = 1$} $\Delta_{c} \Delta_{k} \approx 1.01$. Looking at Tables~\ref{tab:results_LCDM} and \ref{tab:results_CPL}, we can easily check that, from this point of view, the only scenarios which fulfill such condition in an undoubted way are the BM and Moffat models, with a CPL and classical VSL ansatz respectively when a cosmological constant or a  CPL DE model is assumed, i.e. all the models which exhibit the highest favourable evidence. This makes them falsifiable in the next future.

\subsection{Information priors}

The first point to be noted is that when we apply \textit{only} cosmological priors (Tables~\ref{tab:results_LCDM} and \ref{tab:results_CPL}), the information prior condition, $\dot{c}<0$, is automatically verified by the median best fit values for the Avelino \& Martins scenario. Thus, by forcing the condition $\dot{c}<0$ we will have the only consequence to cut the positive-values tail, but not to really improve or worsen the $\chi^2$ value. This can be easily checked in Tables~\ref{tab:results_Info_LCDM} and \ref{tab:results_Info_CPL}, by inspection of the Bayesian Evidence values compared to the reference models, $\Lambda$CDM without and with curvature.

On the other hand, the Barrow \& Magueijo and Moffat scenarios, in the \textit{only} cosmological priors case, always have best fits which satisfy the condition $\dot{c}>0$. Thus, they are intrinsically violating the information prior. As a result, by imposing the information prior, the $\chi^2$ is much more worsened. Actually, while in the cosmological prior case they are the most favoured scenarios, with a Bayesian Evidence even greater than one in some cases, now they have lower values (even turning negative, in the case of a $\Lambda$CDM component) meaning they are clearly disfavoured. And we can only put lower or upper limits on the curvature and on the variation rate of $c(z)$.

\subsection{Entropy priors}

The application of the entropy prior is much more tricky. First of all, from Table~\ref{tab:conditions} one can immediately see that it is impossible to have $\dot{s}>0$, $\dot{c}<0$ and $\rho_{DE}(a)>0$ satisfied at the same time in all cases but the Moffat scenario with a CPL dark energy and a $c$-CPL ansatz for the VSL signal. It must be noted that, with respect to \cite{Chimento2001,Youm2002}, this inconsistency has been obtained not by analytical considerations, but by an extensive scan of the parameters space through Markov Chain Monte Carlo runs. Also, we remember that VSL theories explicitly violate the energy conservation; thus, such inconsistency, is not completely unexpected, and does not mean that the VSL approaches we have considered are wrongly defined, but that a more careful analysis about VSL consequences is needed at every basic level.

As said in previous section, in the Avelino \& Martins case we always have $\dot{c}<0$; this condition is not compatible with $\dot{s}>0$ given in Eq.~(\ref{eq:entropy_AV}). Thus, we have to force $\dot{c}>0$ in order to have $\dot{s}>0$ (see Table~\ref{tab:conditions}). Results are in Tables~\ref{tab:results_Entropy_LCDM} and \ref{tab:results_Entropy_CPL}. As expected, the fit is worst in this case with respect to previous priors cases.

For the Barrow \& Magueijo scenario, we can achieve the condition $\dot{s}>0$ in one case only, with $\dot{c}>0$, see Table \ref{tab:results_Entropy_CPL}. Instead, the only cases within Moffat approach which give reliable and reasonable fits to the data, correspond to the $c$-CPL ansatz, with $n>0$, which actually implies $\dot{c}<0$. Thus, if we had to give full validity to the entropy prior, the Moffat scenario would result to be the only one possible, but with a negative logarithmic Bayesian Evidence ratio, so, it would be disfavoured with respect to the standard $c$-constant $\Lambda$CDM model.

\section{Conclusions}
\label{sec:conclusions}

In this work we have tried to introduce some clarifications in the context of varying speed of light theories. While there is a hard debate about their feasibility, and many theoretical proposals which try to assess all the problems and issues they are generally concerned for, we have detected a global lack of one of the easiest and most natural way to test the reliability of a theory: its application to data, and the resulting statistical analysis. Results in this work are just this: a statistically-based theoretical feasibility study of some VSL approaches; but the way these can be confirmed has to pass, of course, through some dimensionless quantity like, for example, the previously described relative variation of $c$.

Here, we used the most updated cosmological probes and considered three different ways to approach a varying speed of light cosmology. Firstly, we considered the approach initially described in \citep{Albrecht99,Barrow99,BarrowMagueijo99A,BarrowMagueijo99B,BarrowMagueijo99C,BarrowMagueijo00,Magueijo00} (BM) which, even if historically referred as the most influential, is well known to have some theoretical shortcomings as, for example, not a proper way to introduce variation of the speed of light at the action level. At the moment to enforce a comparison with data, this approach looks like a ``mild'' variation to the standard $\Lambda$CDM model, with some consequence only at the continuity equation if the spatial curvature is considered not null. Otherwise, it would be completely equivalent to the standard $c$-constant $\Lambda$CDM scenario. Secondly, we have focussed on a first try to extend such approach, developed by \citep{Avelino99,Avelino00} (AM), and leading to more effective changes to both the Friedmann and the continuity equations, which do not simply disappear even when spatial flatness is assumed. Last, we have paid attention to the newest approach from \citep{Moffat16} (Moffat), which overtakes all the major shortcomings from the first VSL models by considering the varying $c$ as a field, and properly taking it into account since the lagrangian definition. The interesting point is that, while the continuity equation from this model is completely equal to the BM one, the Friedmann equation is different and does not correspond to it even in the limit of null spatial curvature.

One of the most problematic points when facing a VSL theory is the lack of a theoretically-based $c(z)$ ansatz; we have considered two ans\"{a}tze from literature and proposed a new one, linear in the scale factor, which we have called $c$-CPL, in close parallelism with the most used dark energy parametrization. Moreover, we have considered two different cases for a dark energy fluid, i.e. a cosmological constant and a CPL DE parametrization.

Final results show that the most interesting approaches, from a statistical point of view, are the BM and Moffat ones. The AM is able to obtain a quite good fit of present data, but using the Bayesian Evidence as tool to define a preference hierarchy among the models, it results to be the less favourable. On the other hand, the BM approach, while being satisfactory and admissible, has one problem which no statistical probe can quantify: its recognized theoretical unsatisfactory derivation. The most interesting, then, is the VSL approach from Moffat: it has theoretical soundness, and seems to be favoured by data with respect to the stronger and more famous $c$-constant $\Lambda$CDM scenario, both with or without fixing null spatial curvature.
The VSL signal cannot explain present accelerated expansion (one of the historically most required motivations to introduce a varying speed of light) by itself; it always need a dark energy component to enforce it. But there is a $2\sigma$ statistical evidence for both a non-zero VSL signal and for a non-constant DE component. What makes this approach even more interesting is that, while from a pure cosmological point of view, the expected cosmological parameters are quite similar to the $\Lambda$CDM case, there are other observations, like the maximum redshift location in the $D_{A}$, and the amount of signal (as defined in the previous sections) at that redshift, which make it different from the standard scenario and, thus, testable in the very next future.

Finally, we have also considered a VSL analysis by applying priors coming from information theory and the second law of thermodynamics. The application of such priors turned out to be a bit inconclusive and perhaps not completely clear. It seems that there are still some theoretical issues which should be addressed before going to some more general conclusions. Anyway, in this paper we have presented all possible combinations of models and constraints which refer to the topic of varying speed of light cosmologies leaving its deeper attitude to future readers.

{\renewcommand{\tabcolsep}{1.5mm}
{\renewcommand{\arraystretch}{2.}
\begin{table*}[htbp]
\begin{minipage}{\textwidth}
\caption{Scheme showing all the priors that can be applied and if they can effectively give a physical solution or not. A ``\checkmark'' implies that a fit with the assumed conditions is possible, and gives reliable results. A ``$\dag$'' means that the prior conditions can be satisfied by some set of parameters values, but no reliable fit is possible. A ``$\ddag$'' means that there is no set of parameters at all which can satisfy the considered priors.}\label{tab:conditions}
\centering
\resizebox*{0.85\textwidth}{!}{
\begin{tabular}{cccc||cccc}
\hline
\hline
\multicolumn{8}{c}{Cosmological priors} \\
\hline
  id.  & + Information priors & \multicolumn{2}{c||}{+ Entropy priors} & id. & + Information priors & \multicolumn{2}{c}{+ Entropy priors} \\
       &                      & \multicolumn{2}{c||}{$\dot{s}\geq 0 $} &     &                      & \multicolumn{2}{c}{$\dot{s}\geq 0 $} \\
       & $\dot{c}<0$          & $\dot{c}<0$  & $\dot{c}>0$            &     & $\dot{c}<0$          & $\dot{c}<0$  & $\dot{c}>0$ \\
\hline
\multicolumn{8}{c}{Barrow \& Magueijo} \\
$\Lambda$ + $c$-cl.  & \Large{\checkmark} & \Large{\ddag} & \Large{\dag}  & CPL + $c$-cl. & \Large{\checkmark} & \Large{\dag} & \Large{\checkmark} \\
$\Lambda$ + $c$-Mag. & \Large{\checkmark} & \Large{\ddag} & \Large{\dag}  & CPL + $c$-cl. & \Large{\checkmark} & \Large{\ddag} & \Large{\dag}       \\
$\Lambda$ + $c$-CPL  & \Large{\checkmark} & \Large{\dag}  & \Large{\dag}  & CPL + $c$-cl. & \Large{\checkmark} & \Large{\dag}  & \Large{\dag} \\
\hline
\hline
\multicolumn{8}{c}{Avelino \& Martins} \\
$\Lambda$ + $c$-cl.  & \Large{\checkmark} & \Large{\ddag} & \Large{\checkmark} & CPL + $c$-cl. & \Large{\checkmark} & \Large{\ddag} & \Large{\checkmark} \\
$\Lambda$ + $c$-Mag. & \Large{\checkmark} & \Large{\ddag} & \Large{\checkmark} & CPL + $c$-cl. & \Large{\checkmark} & \Large{\ddag} & \Large{\checkmark} \\
$\Lambda$ + $c$-CPL  & \Large{\checkmark} & \Large{\ddag} & \Large{\checkmark} & CPL + $c$-cl. & \Large{\checkmark} & \Large{\ddag} & \Large{\checkmark} \\
\hline
\hline
\multicolumn{8}{c}{Moffat} \\
$\Lambda$ + $c$-cl.  & \Large{\checkmark} & \Large{\ddag} & \Large{\ddag}      & CPL + $c$-cl. & \Large{\checkmark} & \Large{\ddag} & \Large{\dag} \\
$\Lambda$ + $c$-Mag. & \Large{\checkmark} & \Large{\ddag} & \Large{\dag}       & CPL + $c$-cl. & \Large{\checkmark} & \Large{\ddag} & \Large{\dag} \\
$\Lambda$ + $c$-CPL  & \Large{\checkmark} & \Large{\checkmark} & \Large{\ddag} & CPL + $c$-cl. & \Large{\checkmark} & \Large{\checkmark} & \Large{\ddag} \\
\hline
\hline
\end{tabular}}
\end{minipage}
\end{table*}}}

{\renewcommand{\tabcolsep}{1.5mm}
{\renewcommand{\arraystretch}{2.}
\begin{table*}[htbp]
\begin{minipage}{\textwidth}
\caption{Cosmological priors. $\Lambda$CDM case.}\label{tab:results_LCDM}
\centering
\resizebox*{\textwidth}{!}{
\begin{tabular}{c|cccccccccc}
\hline
\hline
  id. & $\Omega_{m}$ & $\Omega_{k}$ & $h$ & $n$ & $\mathcal{B}^{i}_{\Lambda CDM}$ & $\ln \mathcal{B}^{i}_{\Lambda CDM}$ & $z_{M}$ & $\Delta_{c}\Delta_{k}$ & $\Delta_{c}$ & $\Delta_{k}$ \\
\hline
\hline
  & \multicolumn{10}{c}{$\Lambda$CDM} \\
  no $\Omega_{k}$ & $0.309^{+0.005}_{-0.005}$ & $\mathit{0}$ & $0.679^{+0.004}_{-0.004}$ & $-$ & $1$ & $0$ & $1.594^{+0.007}_{-0.007}$ & $\mathit{1}$ & $-$ & $-$ \\
  $\Omega_{k}$ & $0.309^{+0.005}_{-0.005}$ & $0.0008^{+0.0017}_{-0.0016}$ & $0.681^{+0.005}_{-0.005}$ & $-$ & $0.79$ & $-0.24$ & $1.594^{+0.007}_{-0.007}$ & $1.0005^{+0.0010}_{-0.0009}$ & $-$ & $1.0005^{+0.0010}_{-0.0009}$ \\
\hline
\hline
  & \multicolumn{10}{c}{Barrow \& Magueijo} \\
  $c$-cl. & $0.307^{+0.005}_{-0.005}$ & $0.0008^{+0.0010}_{-0.0006}$ & $0.688^{+0.006}_{-0.006}$ & $0.0007^{+0.0005}_{-0.0004}$ & $2.26$ & $0.81$ & $1.594^{+0.007}_{-0.003}$ & $0.9999^{+0.0007}_{-0.0006}$ & $0.9994^{+0.0004}_{-0.0004}$ & $1.0005^{+0.0006}_{-0.0004}$ \\
  $c$-Mag. & $0.307^{+0.005}_{-0.005}$ & $0.0008^{+0.0010}_{-0.0006}$ & $0.689^{+0.006}_{-0.006}$ & $0.0014^{+0.0009}_{-0.0008}$ & $2.40$ & $0.88$ & $1.594^{+0.007}_{-0.003}$ & $0.9999^{+0.0007}_{-0.0006}$ & $0.9993^{+0.0004}_{-0.0004}$ & $1.0005^{+0.0005}_{-0.0004}$ \\
  $c$-CPL & $0.302^{+0.006}_{-0.006}$ & $0.0008^{+0.0009}_{-0.0006}$ & $0.693^{+0.006}_{-0.006}$ & $-0.018^{+0.008}_{-0.008}$ & $12.13$ & $2.50$ & $1.594^{+0.007}_{-0.007}$ & $0.989^{+0.005}_{-0.005}$ & $0.989^{+0.005}_{-0.005}$ & $1.0003^{+0.0003}_{-0.0002}$ \\
\hline
\hline
  & \multicolumn{10}{c}{Avelino \& Martins} \\
  $c$-cl. & $0.322^{+0.014}_{-0.013}$ & $0.006^{+0.006}_{-0.005}$ & $0.683^{+0.005}_{-0.005}$ & $-0.0017^{+0.0016}_{-0.0016}$ & $0.89$ & $-0.11$ & $1.577^{+0.017}_{-0.020}$ & $1.005^{+0.004}_{-0.005}$ & $1.002^{+0.002}_{-0.002}$ & $1.003^{+0.003}_{-0.003}$ \\
  $c$-Mag. & $0.322^{+0.014}_{-0.014}$ & $0.006^{+0.006}_{-0.006}$ & $0.683^{+0.005}_{-0.005}$ & $-0.003^{+0.003}_{-0.003}$ & $0.94$ & $-0.06$ & $1.577^{+0.017}_{-0.020}$ & $1.005^{+0.004}_{-0.004}$ & $1.002^{+0.002}_{-0.002}$ & $1.003^{+0.003}_{-0.003}$ \\
  $c$-CPL & $0.318^{+0.013}_{-0.013}$ & $0.004^{+0.005}_{-0.005}$ & $0.682^{+0.005}_{-0.005}$ & $0.008^{+0.011}_{-0.011}$ & $0.72$ & $-0.33$ & $1.581^{+0.020}_{-0.016}$ & $1.007^{+0.009}_{-0.010}$ & $1.005^{+0.007}_{-0.007}$ & $1.002^{+0.003}_{-0.003}$ \\
\hline
\hline
  & \multicolumn{10}{c}{Moffat} \\
  $c$-cl. & $0.307^{+0.005}_{-0.005}$ & $0.0008^{+0.0011}_{-0.0007}$ & $0.686^{+0.005}_{-0.005}$ & $0.0004^{+0.0004}_{-0.0003}$ & $1.29$ & $0.26$ & $1.597^{+0.007}_{-0.007}$ & $1.0000^{+0.0007}_{-0.0005}$ & $0.9996^{+0.0003}_{-0.0004}$ & $1.0005^{+0.0006}_{-0.0004}$ \\
  $c$-Mag. & $0.307^{+0.005}_{-0.005}$ & $0.0009^{+0.0011}_{-0.0006}$ & $0.686^{+0.005}_{-0.005}$ & $0.0009^{+0.0007}_{-0.0006}$ & $1.27$ & $0.24$ & $1.597^{+0.007}_{-0.007}$ & $1.0000^{+0.0007}_{-0.0005}$ & $0.9996^{+0.0003}_{-0.0004}$ & $1.0005^{+0.0006}_{-0.0004}$ \\
  $c$-CPL & $0.305^{+0.006}_{-0.005}$ & $0.0008^{+0.0010}_{-0.0006}$ & $0.690^{+0.006}_{-0.006}$ & $-0.008^{+0.005}_{-0.005}$ & $3.51$ & $1.26$ & $1.597^{+0.007}_{-0.007}$ & $0.996^{+0.003}_{-0.003}$ & $0.995^{+0.003}_{-0.003}$ & $1.0005^{+0.0005}_{-0.0003}$ \\
\hline
\hline
\end{tabular}}
\end{minipage}
\end{table*}}}

{\renewcommand{\tabcolsep}{1.5mm}
{\renewcommand{\arraystretch}{2.}
\begin{table*}[htbp]
\begin{minipage}{\textwidth}
\caption{Cosmological priors. CPL case.}\label{tab:results_CPL}
\centering
\resizebox*{\textwidth}{!}{
\begin{tabular}{c|cccccccccccc}
\hline
\hline
  id. & $\Omega_{m}$ & $\Omega_{k}$ & $h$ & $n$ & $w_{0}$ & $w_{1}$ & $\mathcal{B}^{i}_{CPL}$ & $\ln \mathcal{B}^{i}_{CPL}$ & $z_{M}$ & $\Delta_{c}\Delta_{k}$ & $\Delta_{c}$ & $\Delta_{k}$ \\
\hline
\hline
  & \multicolumn{12}{c}{CPL} \\
  no $\Omega_{k}$ & $0.302^{+0.006}_{-0.006}$ & $\mathit{0}$ & $0.689^{+0.006}_{-0.006}$ & $-$ & $-1.15^{+0.09}_{-0.08}$ & $0.35^{+0.29}_{-0.32}$ & $1$ & $0$ & $1.584^{+0.013}_{-0.013}$ & $\mathit{1}$ & $-$ & $-$ \\
  $\Omega_{k}$ & $0.302^{+0.006}_{-0.006}$ & $-0.003^{+0.004}_{-0.003}$ & $0.689^{+0.006}_{-0.006}$ & $-$ & $-1.11^{+0.11}_{-0.11}$ & $0.07^{+0.52}_{-0.61}$ & $0.75$ & $-0.29$ & $1.594^{+0.017}_{-0.017}$ & $0.998^{+0.002}_{-0.002}$ & $-$ & $0.998^{+0.002}_{-0.002}$ \\
\hline
\hline
  & \multicolumn{12}{c}{Barrow \& Magueijo} \\
  $c$-cl. & $0.301^{+0.006}_{-0.005}$ & $-0.002^{+0.004}_{-0.004}$ & $0.695^{+0.007}_{-0.007}$ & $0.003^{+0.002}_{-0.002}$ & $-1.14^{+0.08}_{-0.08}$ & $0.74^{+0.20}_{-0.17}$ & $4.04$ & $1.40$ & $1.564^{+0.010}_{-0.010}$ & $0.997^{+0.003}_{-0.004}$ & $0.997^{+0.002}_{-0.002}$ & $0.999^{+0.002}_{-0.002}$ \\
  $c$-Mag. & $0.301^{+0.006}_{-0.005}$ & $-0.005^{+0.005}_{-0.004}$ & $0.696^{+0.006}_{-0.007}$ & $0.008^{+0.003}_{-0.003}$ & $-1.09^{+0.08}_{-0.08}$ & $0.57^{+0.23}_{-0.16}$ & $8.44$ & $2.13$ & $-$ & $-$ & $-$ & $-$ \\
  $c$-CPL & $0.296^{+0.006}_{-0.006}$ & $0.001^{+0.004}_{-0.003}$ & $0.696^{+0.006}_{-0.007}$ & $-0.031^{+0.016}_{-0.016}$ & $-1.14^{+0.07}_{-0.08}$ & $0.64^{+0.24}_{-0.16}$ & $3.55$ & $1.27$ & $1.561^{+0.010}_{-0.013}$ & $0.982^{+0.010}_{-0.010}$ & $0.981^{+0.010}_{-0.009}$ & $1.001^{+0.002}_{-0.002}$ \\
\hline
\hline
  & \multicolumn{12}{c}{Avelino \& Martins} \\
  $c$-cl. & $0.324^{+0.016}_{-0.014}$ & $0.004^{+0.006}_{-0.005}$ & $0.693^{+0.006}_{-0.006}$ & $-0.003^{+0.002}_{-0.002}$ & $-1.05^{+0.13}_{-0.11}$ & $-0.38^{+0.62}_{-0.79}$ & $1.27$ & $0.24$ & $1.574^{+0.020}_{-0.020}$ & $1.005^{+0.005}_{-0.004}$ & $1.003^{+0.002}_{-0.002}$ & $1.003^{+0.003}_{-0.003}$ \\
  $c$-Mag. & $0.325^{+0.015}_{-0.014}$ & $0.005^{+0.006}_{-0.005}$ & $0.693^{+0.006}_{-0.007}$ & $-0.006^{+0.003}_{-0.003}$ & $-1.05^{+0.13}_{-0.12}$ & $-0.38^{+0.66}_{-0.78}$ & $1.36$ & $0.31$ & $1.574^{+0.020}_{-0.020}$ & $1.006^{+0.004}_{-0.004}$ & $1.003^{+0.001}_{-0.002}$ & $1.003^{+0.003}_{-0.003}$ \\
  $c$-CPL & $0.325^{+0.016}_{-0.015}$ & $0.004^{+0.005}_{-0.005}$ & $0.693^{+0.006}_{-0.007}$ & $0.021^{+0.012}_{-0.012}$ & $-1.05^{+0.14}_{-0.12}$ & $-0.44^{+0.66}_{-0.86}$ & $1.38$ & $0.32$ & $1.574^{+0.020}_{-0.020}$ & $1.016^{+0.009}_{-0.010}$ & $1.013^{+0.007}_{-0.007}$ & $1.002^{+0.003}_{-0.003}$ \\
\hline
\hline
  & \multicolumn{12}{c}{Moffat} \\
  $c$-cl. & $0.288^{+0.010}_{-0.013}$ & $-0.011^{+0.010}_{-0.011}$ & $0.694^{+0.006}_{-0.006}$ & $0.006^{+0.004}_{-0.003}$ & $-1.08^{+0.09}_{-0.09}$ & $0.72^{+0.09}_{-0.09}$ & $9.87$ & $2.29$ & $1.584^{+0.020}_{-0.017}$ & $0.990^{+0.008}_{-0.008}$ & $0.995^{+0.003}_{-0.003}$ & $0.995^{+0.005}_{-0.005}$ \\
  $c$-Mag. & $0.301^{+0.006}_{-0.006}$ & $0.001^{+0.003}_{-0.003}$ & $0.691^{+0.006}_{-0.006}$ & $0.002^{+0.003}_{-0.003}$ & $-1.15^{+0.10}_{-0.07}$ & $0.71^{+0.32}_{-0.97}$ & $0.35$ & $0.78$ & $1.577^{+0.023}_{-0.023}$ & $0.9997^{+0.0008}_{-0.0015}$ & $1.000^{+0.001}_{-0.002}$ & $1.0000^{+0.0001}_{-0.0002}$ \\
  $c$-CPL & $0.296^{+0.006}_{-0.006}$ & $-0.001^{+0.004}_{-0.004}$ & $0.696^{+0.007}_{-0.007}$ & $-0.033^{+0.016}_{-0.015}$ & $-1.10^{+0.07}_{-0.08}$ & $0.59^{+0.21}_{-0.14}$ & $6.42$ & $1.86$ & $1.574^{+0.010}_{-0.016}$ & $1.016^{+0.021}_{-0.022}$ & $0.979^{+0.009}_{-0.009}$ & $1.038^{+0.016}_{-0.016}$ \\
\hline
\hline
\end{tabular}}
\end{minipage}
\end{table*}}}

{\renewcommand{\tabcolsep}{1.5mm}
{\renewcommand{\arraystretch}{2.}
\begin{table*}[htbp]
\begin{minipage}{\textwidth}
\caption{Cosmological plus information priors. $\Lambda$CDM case.}\label{tab:results_Info_LCDM}
\centering
\resizebox*{0.6\textwidth}{!}{
\begin{tabular}{c|cccccc}
\hline
\hline
  id. & $\Omega_{m}$ & $\Omega_{k}$ & $h$ & $n$ & $\mathcal{B}^{i}_{\Lambda CDM}$ & $\ln \mathcal{B}^{i}_{\Lambda CDM}$ \\
\hline
\hline
  & \multicolumn{6}{c}{$\Lambda$CDM} \\
  no $\Omega_{k}$ & $0.309^{+0.005}_{-0.005}$ & $\mathit{0}$ & $0.679^{+0.004}_{-0.004}$ & $-$ & $1$ & $0$ \\
  $\Omega_{k}$ & $0.309^{+0.005}_{-0.005}$ & $0.0008^{+0.0017}_{-0.0016}$ & $0.681^{+0.005}_{-0.005}$ & $-$ & $0.79$ & $-0.24$ \\
\hline
\hline
  & \multicolumn{6}{c}{Barrow \& Magueijo} \\
  $c$-cl. & $0.311^{+0.005}_{-0.005}$ & $>-5\cdot10^{-4}$ & $0.676^{+0.004}_{-0.004}$ & $>-9\cdot 10^{-5}$ & $0.29$ & $-1.22$ \\
  $c$-Mag. & $0.310^{+0.005}_{-0.005}$ & $>-5\cdot10^{-4}$ & $0.676^{+0.004}_{-0.004}$ & $>-2\cdot 10^{-4}$ & $0.30$ & $-1.21$ \\
  $c$-CPL & $0.311^{+0.005}_{-0.005}$ & $>-6\cdot10^{-4}$ & $0.676^{+0.006}_{-0.006}$ & $<0.003$ & $0.28$ & $-1.29$ \\
\hline
\hline
  & \multicolumn{6}{c}{Avelino \& Martins} \\
  $c$-cl. & $0.326^{+0.012}_{-0.010}$ & $0.008^{+0.005}_{-0.004}$ & $0.684^{+0.005}_{-0.005}$ & $-0.0022^{+0.0012}_{-0.0013}$ & $1.00$ & $-0.003$ \\
  $c$-Mag. & $0.326^{+0.012}_{-0.010}$ & $0.008^{+0.005}_{-0.004}$ & $0.684^{+0.005}_{-0.005}$ & $-0.0043^{+0.0023}_{-0.0027}$ & $1.00$ & $-0.004$ \\
  $c$-CPL & $0.323^{+0.011}_{-0.009}$ & $0.006^{+0.004}_{-0.004}$ & $0.683^{+0.005}_{-0.005}$ & $0.012^{+0.009}_{-0.007}$ & $0.79$ & $-0.24$ \\
\hline
\hline
  & \multicolumn{6}{c}{Moffat} \\
  $c$-cl. & $0.311^{+0.005}_{-0.005}$ & $>-6\cdot10^{-4}$ & $0.676^{+0.004}_{-0.004}$ & $>-8\cdot10^{-5}$ & $0.30$ & $-1.21$ \\
  $c$-Mag. & $0.311^{+0.005}_{-0.005}$ & $>-6\cdot10^{-4}$ & $0.676^{+0.004}_{-0.004}$ & $>-2\cdot10^{-4}$ & $0.27$ & $-1.32$ \\
  $c$-CPL & $0.311^{+0.005}_{-0.005}$ & $>-6\cdot10^{-4}$ & $0.676^{+0.004}_{-0.004}$ & $<0.002$ & $0.29$ & $-1.23$ \\
\hline
\hline
\end{tabular}}
\end{minipage}
\end{table*}}}

{\renewcommand{\tabcolsep}{1.5mm}
{\renewcommand{\arraystretch}{2.}
\begin{table*}[htbp]
\begin{minipage}{\textwidth}
\caption{Cosmological plus information priors. CPL case.}\label{tab:results_Info_CPL}
\centering
\resizebox*{0.75\textwidth}{!}{
\begin{tabular}{c|cccccccc}
\hline
\hline
  id. & $\Omega_{m}$ & $\Omega_{k}$ & $h$ & $n$ & $w_{0}$ & $w_{1}$ & $\mathcal{B}^{i}_{CPL}$ & $\ln \mathcal{B}^{i}_{CPL}$ \\
\hline
\hline
  & \multicolumn{8}{c}{CPL} \\
  no $\Omega_{k}$ & $0.302^{+0.006}_{-0.006}$ & $\mathit{0}$ & $0.689^{+0.006}_{-0.006}$ & $-$ & $-1.15^{+0.09}_{-0.08}$ & $0.35^{+0.29}_{-0.32}$ & $1$ & $0$ \\
  $\Omega_{k}$ & $0.302^{+0.006}_{-0.006}$ & $-0.003^{+0.004}_{-0.003}$ & $0.689^{+0.006}_{-0.006}$ & $-$ & $-1.11^{+0.11}_{-0.11}$ & $0.07^{+0.52}_{-0.61}$ & $0.75$ & $-0.29$ \\
\hline
\hline
  & \multicolumn{8}{c}{Barrow \& Magueijo} \\
  $c$-cl. & $0.302^{+0.006}_{-0.006}$ & $-0.002^{+0.003}_{-0.003}$ & $0.688^{+0.006}_{-0.006}$ & $>-2\cdot10^{-4}$ & $-1.12^{+0.11}_{-0.12}$ & $-0.02^{+0.70}_{-0.63}$ & $0.38$ & $-0.95$ \\
  $c$-Mag. & $0.302^{+0.006}_{-0.006}$ & $-0.002^{+0.005}_{-0.003}$ & $0.688^{+0.006}_{-0.006}$ & $>-5\cdot10^{-4}$ & $-1.14^{+0.13}_{-0.12}$ & $0.07^{+0.67}_{-0.68}$ & $0.36$ & $-1.02$ \\
  $c$-CPL & $0.304^{+0.007}_{-0.006}$ & $0.003^{+0.007}_{-0.004}$ & $0.689^{+0.007}_{-0.006}$ & $<0.009$ & $-1.27^{+0.08}_{-0.08}$ & $0.86^{+0.40}_{-0.28}$ & $0.25$ & $-1.37$ \\
\hline
\hline
  & \multicolumn{8}{c}{Avelino \& Martins} \\
  $c$-cl. & $0.326^{+0.014}_{-0.013}$ & $0.005^{+0.005}_{-0.005}$ & $0.693^{+0.006}_{-0.006}$ & $-0.003^{+0.001}_{-0.002}$ & $-1.04^{+0.13}_{-0.12}$ & $-0.44^{+0.65}_{-0.82}$ & $1.44$ & $0.36$ \\
  $c$-Mag. & $0.326^{+0.014}_{-0.013}$ & $0.005^{+0.005}_{-0.005}$ & $0.693^{+0.006}_{-0.006}$ & $-0.006^{+0.003}_{-0.003}$ & $-1.05^{+0.13}_{-0.12}$ & $-0.38^{+0.66}_{-0.78}$ & $1.39$ & $0.33$ \\
  $c$-CPL & $0.326^{+0.015}_{-0.013}$ & $0.004^{+0.005}_{-0.005}$ & $0.693^{+0.006}_{-0.006}$ & $0.022^{+0.011}_{-0.010}$ & $-1.04^{+0.14}_{-0.12}$ & $-0.54^{+0.67}_{-0.84}$ & $1.65$ & $0.50$ \\
\hline
\hline
  & \multicolumn{8}{c}{Moffat} \\
  $c$-cl. & $0.303^{+0.006}_{-0.006}$ & $>-3\cdot10^{-4}$ & $0.688^{+0.006}_{-0.006}$ & $0.006^{+0.004}_{-0.003}$ & $-1.08^{+0.11}_{-0.10}$ & $-0.21^{+0.47}_{-0.57}$ & $0.43$ & $-0.85$ \\
  $c$-Mag. & $0.303^{+0.006}_{-0.006}$ & $-0.003^{+0.002}_{-0.002}$ & $0.688^{+0.006}_{-0.006}$ & $>-6\cdot10^{-4}$ & $-1.08^{+0.10}_{-0.09}$ & $-0.23^{+0.42}_{-0.52}$ & $0.42$ & $-0.86$ \\
  $c$-CPL & $0.303^{+0.006}_{-0.006}$ & $0.003^{+0.004}_{-0.003}$ & $0.689^{+0.006}_{-0.006}$ & $<0.007$ & $-1.27^{+0.09}_{-0.10}$ & $0.81^{+0.36}_{-0.25}$ & $0.31$ & $-1.17$ \\
\hline
\hline
\end{tabular}}
\end{minipage}
\end{table*}}}

{\renewcommand{\tabcolsep}{1.5mm}
{\renewcommand{\arraystretch}{2.}
\begin{table*}[htbp]
\begin{minipage}{\textwidth}
\caption{Cosmological plus entropy priors. $\Lambda$CDM case.}\label{tab:results_Entropy_LCDM}
\centering
\resizebox*{0.65\textwidth}{!}{
\begin{tabular}{c|cccccc}
\hline
\hline
  id. & $\Omega_{m}$ & $\Omega_{k}$ & $h$ & $n$ & $\mathcal{B}^{i}_{\Lambda CDM}$ & $\ln \mathcal{B}^{i}_{\Lambda CDM}$ \\
\hline
\hline
  & \multicolumn{6}{c}{$\Lambda$CDM} \\
  no $\Omega_{k}$ & $0.309^{+0.005}_{-0.005}$ & $\mathit{0}$ & $0.679^{+0.004}_{-0.004}$ & $-$ & $1$ & $0$ \\
  $\Omega_{k}$ & $0.309^{+0.005}_{-0.005}$ & $0.0008^{+0.0017}_{-0.0016}$ & $0.681^{+0.005}_{-0.005}$ & $-$ & $0.79$ & $-0.24$ \\
\hline
\hline
  & \multicolumn{6}{c}{Avelino \& Martins} \\
  $c$-cl. & $0.303^{+0.007}_{-0.008}$ & $-0.002^{+0.002}_{-0.003}$ & $0.680^{+0.005}_{-0.005}$ & $<1.2\cdot10^{-3}$ & $0.42$ & $-0.88$ \\
  $c$-Mag. & $0.303^{+0.007}_{-0.008}$ & $-0.002^{+0.002}_{-0.003}$ & $0.679^{+0.005}_{-0.005}$ & $<2.4\cdot10^{-3}$ & $0.43$ & $-0.86$ \\
  $c$-CPL & $0.302^{+0.007}_{-0.008}$ & $-0.002^{+0.002}_{-0.003}$ & $0.680^{+0.005}_{-0.005}$ & $>-3.7\cdot10^{-3}$ & $0.45$ & $-0.81$ \\
\hline
\hline
  & \multicolumn{6}{c}{Moffat} \\
  $c$-CPL & $0.304^{+0.006}_{-0.006}$ & $>-6.7\cdot10^{-4}$ & $0.684^{+0.006}_{-0.006}$ & $0.003^{+0.001}_{-0.001}$ & $3.28$ & $1.19$ \\
\hline
\hline
\end{tabular}}
\end{minipage}
\end{table*}}}

{\renewcommand{\tabcolsep}{1.5mm}
{\renewcommand{\arraystretch}{2.}
\begin{table*}[htbp]
\begin{minipage}{\textwidth}
\caption{Cosmological plus entropy priors. CPL case.}\label{tab:results_Entropy_CPL}
\centering
\resizebox*{0.75\textwidth}{!}{
\begin{tabular}{c|cccccccc}
\hline
\hline
  id. & $\Omega_{m}$ & $\Omega_{k}$ & $h$ & $n$ & $w_{0}$ & $w_{1}$ & $\mathcal{B}^{i}_{CPL}$ & $\ln \mathcal{B}^{i}_{CPL}$ \\
\hline
\hline
  & \multicolumn{8}{c}{CPL} \\
  no $\Omega_{k}$ & $0.302^{+0.006}_{-0.006}$ & $\mathit{0}$ & $0.689^{+0.006}_{-0.006}$ & $-$ & $-1.15^{+0.09}_{-0.08}$ & $0.35^{+0.29}_{-0.32}$ & $1$ & $0$ \\
  $\Omega_{k}$ & $0.302^{+0.006}_{-0.006}$ & $-0.003^{+0.004}_{-0.003}$ & $0.689^{+0.006}_{-0.006}$ & $-$ & $-1.11^{+0.11}_{-0.11}$ & $0.07^{+0.52}_{-0.61}$ & $0.75$ & $-0.29$ \\
\hline
\hline
  & \multicolumn{8}{c}{Barrow \& Magueijo} \\
  $c$-cl. & $0.303^{+0.006}_{-0.005}$ & $-0.004^{+0.003}_{-0.004}$ & $0.697^{+0.006}_{-0.007}$ & $0.004^{+0.002}_{-0.001}$ & $-1.11^{+0.07}_{-0.08}$ & $0.71^{+0.15}_{-0.12}$ & $5.43$ & $1.69$ \\
\hline
\hline
  & \multicolumn{8}{c}{Avelino \& Martins} \\
  $c$-cl. & $0.296^{+0.008}_{-0.008}$ & $-0.004^{+0.004}_{-0.004}$ & $0.687^{+0.006}_{-0.006}$ & $<0.001$ & $-1.13^{+0.11}_{-0.09}$ & $0.21^{+0.46}_{-0.57}$ & $0.37$ & $-1.01$ \\
  $c$-Mag. & $0.296^{+0.007}_{-0.008}$ & $-0.004^{+0.005}_{-0.004}$ & $0.688^{+0.006}_{-0.006}$ & $<0.002$ & $-1.13^{+0.11}_{-0.10}$ & $0.19^{+0.53}_{-0.62}$ & $0.39$ & $-0.95$ \\
  $c$-CPL & $0.297^{+0.007}_{-0.008}$ & $-0.004^{+0.004}_{-0.004}$ & $0.688^{+0.006}_{-0.006}$ & $>-0.003$ & $-1.12^{+0.10}_{-0.10}$ & $0.22^{+0.47}_{-0.56}$ & $0.43$ & $-0.84$ \\
\hline
\hline
  & \multicolumn{8}{c}{Moffat} \\
  $c$-CPL & $0.302^{+0.006}_{-0.005}$ & $>-0.001$ & $0.689^{+0.006}_{-0.006}$ & $<0.002$ & $-1.12^{+0.18}_{-0.17}$ & $0.15^{+0.58}_{-0.81}$ & $0.72$ & $-0.33$ \\
\hline
\hline
\end{tabular}}
\end{minipage}
\end{table*}}}

\section*{Acknowledgements}

The authors warmly thank John Moffat for the discussions related to the topic and for reading a preliminary version of the paper. This work was financed by the Polish National Science Center Grant DEC-2012/06/A/ST2/00395. The use of the CI\'S computer cluster at the National Centre for Nuclear Research is gratefully acknowledged.

\vfill
\end{document}